# Heavy ion acceleration using femtosecond laser pulses

G. M. Petrov[1], C. McGuffey[2], A. G. R. Thomas[3], K. Krushelnick[3] and F. N. Beg[2]

[1]Naval Research Laboratory, Plasma Physics Division, 4555 Overlook Ave. SW, Washington, DC 20375, USA

[2]Mechanical and Aerospace Engineering and Center for Energy Research, University of California-San Diego, La Jolla, CA 92093 USA

[3]Center for Ultrafast Optical Science, University of Michigan, Ann Arbor, MI 48109 USA

e-mail: george.petrov@nrl.navy.mil

**Abstract**

Theoretical study of heavy ion acceleration from ultrathin (<200 nm) gold foils irradiated by a short pulse laser is presented. Using two dimensional particle-in-cell simulations the time history of the laser bullet is examined in order to get insight into the laser energy deposition and ion acceleration process. For laser pulses with intensity $3 \times 10^{21}$ W/cm$^2$, duration 32 fs, focal spot size 5 µm and energy 27 Joules the calculated reflection, transmission and coupling coefficients from a 20 nm foil are 80 %, 5 % and 15 %, respectively. The conversion efficiency into gold ions is 8 %. Two highly collimated counter-propagating ion beams have been identified. The forward accelerated gold ions have average and maximum charge-to-mass ratio of 0.25 and 0.3, respectively, maximum normalized energy 25 MeV/nucleon and flux $2 \times 10^{11}$ ions/sr. Analytical model was used to determine a range of foil thicknesses suitable for acceleration of gold ions in the Radiation Pressure Acceleration regime and the onset of the Target Normal Sheath Acceleration regime. The numerical simulations and analytical model point to at least four technical challenges hindering the heavy ion acceleration: low charge-to-mass ratio, limited number of ions amenable to acceleration, delayed acceleration and high reflectivity of the plasma. Finally, a regime suitable for heavy ion acceleration has been identified in an alternative approach by analyzing the energy absorption and distribution among participating species and scaling of conversion efficiency, maximum energy, and flux with laser intensity.

## 1. Introduction

Short pulse lasers have been extensively used for generation of intense multi-MeV ion beams. For many years the increase in maximum energy and conversion efficiency into ions has been incremental. But the laser technology and target preparation have experienced marked improvement, setting the stage for a leap in laser-driven ion acceleration. Clean laser pulses with intensity $I > 10^{21} W/cm^2$ and ultrathin (nm) targets are now available and have been used in a number of experiments, making long-standing predictions of advanced acceleration schemes a reality. The Target Normal Sheath Acceleration (TNSA) [1] has been the hallmark of ion acceleration for nearly two decades, but it is now possible to go beyond TNSA and reach more favorable regimes such as Radiation Pressure Acceleration (RPA) in circular [2,3,4,*5*] and linear [6,7] polarizations, Breakout Afterburner (BoA) [8], "laser-piston" (LP) [9,10] and Relativistic Induced Transparency (RIT) [11]. Transition from TNSA to RPA [12] and BoA [13,14] for protons and carbon ions has been experimentally demonstrated along with other impressive results: 40 MeV protons from a laser system with only 7.5 Joules of laser energy on target [15], $C_6^+$ ions with energies exceeding 80 MeV/nucleon [14] and ~1 GeV fully stripped Fe ions [16], to name a few. In all these studies the focus was on protons and light ions, for which the above mentioned acceleration mechanisms have been attributed. Mid-Z ions were also investigated [16,17,18], while for heavy ions only a handful of experimental [19,20] and theoretical studies [21,22] exist. No acceleration mechanisms have been identified for mid- and high-Z ions. Braenzel *et al* [20] developed an analytical model to elucidate the steep dependence of the maximum energy of gold ions as a function of ion charge, but the exact acceleration process remains unknown. The matter is even more complicated since these ions can originate from different parts of the target: bulk [17,18] or from a thin layer on the rear surface of the foil, akin to contaminants [16]. This implies that different acceleration mechanisms can be at play depending on the location of ions of interest in the target (bulk or surface). With plethora of experimental and theoretical studies devoted to protons and light ions, the next logical step is to extend the research to the more challenging case of heavy ions such as Au or W. It is of fundamental interest to understand the intricate details and issues relevant to heavy ion beam acceleration. The present study has been motivated by three factors, which come together as a results of recent breakthroughs in laser development and theoretical advancements in the field: (i) the issues and physics of heavy ion acceleration are unknown; (ii) the laser and target parameter landscape has not been mapped, e.g. it is unknown what combination of laser systems and targets will work best; and (iii) the availability of ultra-high contrast lasers (>$10^{10}$) and ultrathin foils (down to 5-10 nm), which allows the exploration of a wide variety of ion acceleration mechanisms.

The paper is organized into analytical part (Sections 2 and 3) and modeling and simulations part (Sections 4, 5 and 6). In Section 2 we review the requirements and challenges facing the acceleration of heavy ions. In Section 3 we map the ion acceleration mechanisms versus foil thickness. An example of heavy ion acceleration in the RPA regime is presented in Section 4, where numerical simulations for gold ion acceleration from sub-micron foils are carried out using a 2D3V particle-in-cell (PIC) code. Analogous results in the TNSA regime are presented in Section 5. In Section 6 the ion acceleration is analyzed in terms of energy absorption and partition. The conversion efficiency scaling of gold and contaminants ions with laser intensity is investigated. A summary of the results is given in the final section of the paper.

## 2. Challenges for heavy ion acceleration

Before we go into details of the heavy ion acceleration process, we first need to outline the relevant issues, as well as the conditions appropriate for acceleration of heavy ions. There are numerous differences compared to ions from low-Z material:

- Lower charge-to-mass ratio: For heavy ions, *q/M* is twice lower compared to light ions, which has implications for the maximum ion energy the heavy ions can reach, as well as the competition between heavy ions and the ever-present contaminants on the target surface.

- Fewer ions available for acceleration: Only those in the focal spot having large *q/M* can be efficiently accelerated.

- Delayed acceleration: Heavy ions can be accelerated only after the peak of the laser pulse.

- Plasma mirror effect: Due to the large ion charge in the focal spot ($q \cong 50$), the electron density becomes extremely high, exceeding 2000 times the critical electron density right at the moment the ion acceleration starts.

Unlike protons and light ions, the acceleration of heavy ions is plagued with problems. The most critical one is the low charge-to-mass ratio *q/M*, since the normalized ion energy scales as $E/M \sim (q/M)^2$ [20,23,24]. For gold the estimated maximum ion charge and charge-to-mass ratio are $q \cong 70$ and $(q/M)_{max} \cong 0.35$, respectively. This is illustrated in Figure 1, which plots the maximum charge-to-mass ratio versus laser intensity. It is based on the so-called Bethe rule, $I_{th} = \frac{2.2 \times 10^{15}}{\overline{z}^2} \left( \frac{I_p(\overline{z})}{27.21} \right)^2$, which relates the threshold laser intensity $I_{th}$ (in units [W/cm$^2$]) for optical field ionization to the maximum reachable charge $\overline{q} = \overline{z} - 1$ of ion having ionization potential $I_p(\overline{z})$ (in units [eV]) [25,26]. In reality, the most likely ion charge and charge-to-mass ratio for Au ions are $q \cong 50$ and $q/M \cong 0.25$, respectively, as it will be shown later with simulations. From Figure 1 we conclude that laser intensities below about $I \sim 10^{20} W/cm^2$ are inadequate for heavy ion acceleration. The low *q/M* is disadvantageous for Au and entails the well-known "contaminants problem": a thin (2-3 nm) layer of hydrocarbon or water residing on the surface of the foil steals nearly all the energy coupled to the plasma and suppresses the acceleration of heavy ions. This effect has already been seen experimentally for mid-Z ions [17,18].

The second problem is the limited number of heavy ions that can be accelerated. Since *q/M* is very sensitive to *I*, the only useful ions amenable to acceleration reside in the laser focal spot. This is in contrast to low-Z ions, e.g. carbon, which can be fully ionized at much lower intensities and the available ions for acceleration extend into the wings of the laser intensity profile. Since the number of atoms in the foil scales with distance from focal spot center as $r^2$, we estimate that the number of gold ions that can be efficiently accelerated is at least one order of magnitude less than the corresponding number of carbon ions and protons.

The third issue is more subtle and is unique for heavy ions. Ionization and acceleration are divided into two distinct phases separated in time, e.g. a phase of ionization and a phase of acceleration. During the first phase the ions must be ionized to very high charge states, a process that completes at the peak of the laser pulse. The second phase, acceleration, takes place during the pulse fall-off, shortening the time available for acceleration by a factor of two. The two phases are shown in Figure 2 aided by 2D PIC simulations. The laser pulses and foil parameters are listed in Table 1. The maximum ion energy and conversion efficiency into gold ions increase sharply, but only after the peak of the laser pulse. The ion acceleration is "delayed" until the laser pulse reaches its peak, and only half of the pulse can be used to accelerate ions, which may prevent ions from reaching full velocity. Thus short laser pulses (30-40 fs), which are attractive for acceleration of light ions, are borderline adequate for heavy ions due to insufficient acceleration time. This drawback can be compensated by increasing the laser intensity, which once again leads to the conclusion that high intensities are required.

The fourth and final problem is the reflectivity of the target. Ion acceleration commences at the peak of the laser pulse, when the ion charge is already high (Figure 2). The electron density reaches values on the order of $n_e \cong \bar{q} n_{Au} \cong 3 \times 10^{24} cm^{-3}$, which results in a plasma that is $n_e/n_{cr} > 2000$ times overdense. The "plasma mirror" reflects most of the incoming laser radiation (cf. Figure 8), reducing coupling of laser energy to ions. All these issues adversely affect the formation of heavy ion beams. In the following sections we will suggest approaches to overcome them.

## 3. Acceleration mechanisms for heavy ion beams

One of the advanced acceleration schemes exhibiting superior scaling is RPA. For RPA to work, the target must remain overdense for the duration of the pulse. In addition, in the "thin foil" regime, the hole-boring process must reach the rear of the foil before the laser pulse ends, which imposes limitations on the foil thickness. We selected a range of peak laser intensities, $3 \times 10^{20} < I_0 < 3 \times 10^{21} W/cm^2$, suitable for RPA and in accordance with Figure 1. We will begin by introducing a useful scale length and relate other parameters such as foil thickness to it. Perhaps the most important one is the relativistic skin depth $\ell_{skin} = \gamma^{1/2} c / \omega_p$, where $\gamma$ and $\omega_p$ are the relativistic parameter and electron plasma frequency, and $c$ is the speed of light. For simplicity $\gamma$ and $\omega_p$ are taken at the peak of the laser pulse. The skin depth is chosen because it is convenient (comparable to foil thickness), separates "transparent" from "opaque" foils and the energy absorption reaches maximum for foil thickness comparable to the skin depth. After a few simple manipulations the skin depth takes form

$$\ell_{skin} = \left( \frac{\gamma n_{cr}}{n_e} \right)^{1/2} \frac{\lambda_0}{2\pi}. \tag{1}$$

The right hand side scales weakly with laser intensity, $\ell_{skin} \sim \gamma^{1/2} \cong a_0^{1/2} \sim I_0^{1/4}$. For typical laser and plasma parameters in the focal spot, $\gamma \cong 12 - 37$, $\bar{q} \cong 50$, electron density $n_e \cong \bar{q} n_{Au} \cong 3 \times 10^{24} cm^{-3}$, and critical density $n_{cr} \cong 1.8 \times 10^{21} cm^{-3}$, Formula (1) yields $\ell_{skin} \cong (0.014 - 0.024)\lambda_0 \cong 11 - 20 nm$.

The second parameter of importance for RPA is the optimal foil thickness $\ell^{opt}$ derived from the condition $\frac{n_e}{n_{cr}} \frac{\ell^{opt}}{\lambda_0} \cong a_0$ [4,6,27,28], stating that the normalized areal density is equal to the normalized laser field amplitude $a_0 = 8.5 \times 10^{-10} \sqrt{I_0} \lambda_0$. In the above formulas $I_0$ is in units W/cm$^2$ and $\lambda_0$ is in units of μm. Using Formula (1), it can be written in an alternative form,

$$\frac{\ell^{opt}}{\ell_{skin}} \cong \frac{4\pi \ell_{skin}}{\lambda_0}. \tag{2}$$

For high-Z material the right hand side is between 0.2 and 0.3, e. g. the optimum foil thickness for gold in the RPA regime is 1/4 of the relativistic skin depth. For foil thickness $L \leq \ell^{opt}$ the RPA is unstable with all electrons blown out of the foil. For stable RPA, the foil thickness must be larger than the "optimal thickness" given by Equation (2).

The third scale length of importance is the hole-boring length $\ell_{HB} = v_{HB}\tau_{HB}$, which divides the RPA into hole-boring (HB) and light sail (LS) regimes [27,28]. The normalized hole-boring velocity, $\frac{v_{HB}}{c} = \sqrt{\frac{\bar{q}}{M}\frac{m_e}{m_p}\frac{n_{cr}}{n_e}}a_0$ [27,28], is the recession velocity of the plasma surface driven by the laser piston. The difference between light and heavy ions becomes immediately apparent considering the scaling with ion mass, $v_{HB} \sim M^{-1/2}$. The hole-boring velocity for gold is four times slower compared to that of carbon. For the hole boring time we can take the ion acceleration time, e.g. $\tau_{HB} \cong \tau_{FWHM}$. In order to accelerate ions in the RPA-LS regime, $L < \ell_{HB}$ is required. Using again the expression for the skin depth (1), the hole-boring length can be written as $\ell_{HB} = \sqrt{\frac{\bar{q}}{M}\frac{m_e}{m_p}}2\pi N_{laser}a_0^{1/2}\ell_{skin}$. Combining the two conditions, $L > \ell^{opt}$ for stable RPA, and $L < \ell_{HB}$ for LS-RPA, we arrive at:

$$\frac{4\pi\ell_{skin}}{\lambda_0} < \frac{L}{\ell_{skin}} < \sqrt{\frac{\bar{q}}{M}\frac{m_e}{m_p}}2\pi N_{laser}a_0^{1/2}. \qquad (3)$$

The right hand side of (3), assuming charge-to-mass ratio $\frac{\bar{q}}{M} \cong \frac{1}{4}$, number of laser periods $N_{laser} = \frac{c\tau_{FWHM}}{\lambda_0} = 12$ and $a_0 \cong 12 - 37$, is between 4 and 6. Thus for a typical short pulse laser (30-40 fs) the foil thickness in the RPA-LS regime is limited in the interval

$$\frac{1}{4}\ell_{skin} < L < 5\ell_{skin}. \qquad (4)$$

In absolute units, it is between 5 and 100 nm. Equation (4) is simple and has a clear physical meaning: RPA-LS is realized for foil thickness comparable to the skin depth. All ions in the focal spot can be volumetrically accelerated, which is very efficient and optimizes the energy absorption [8,29]. Another advantage of using Formulas (3) or (4) is that both sides scale weakly with laser intensity, $\sim I_0^{1/4}$.

The regime landscape is illustrated schematically in Fig. 3. Acceleration from foils that are too thin and $L < \frac{1}{4}\ell_{skin}$ holds, is inherently unstable and corresponds to the Coulomb Explosion (CE) regime. For foil thickness obeying $\frac{1}{4}\ell_{skin} < L < 5\ell_{skin}$ the ion acceleration is formally in the RPA-LS regime, and for $L > 5\ell_{skin}$ the conventional TNSA takes place. For full dominance of RPA over TNSA it is also required that the maximum velocity of the ions (about twice the hole-boring velocity) exceeds the maximum ion velocity obtained by TNSA [6].

**4. Heavy ion acceleration in the RPA regime**

Numerical simulations for gold ion acceleration in the RPA regime are performed using a two-dimensional electromagnetic PIC code [30,31]. The target is a flat 20 nm Au foil covered with a 5 nm contaminant layer residing on the back of the foil, located at spatial position $x = 48\,\mu m$. For numerical purposes, the contaminants are modeled as a thin sheet of water at liquid density. The foil thickness is chosen to roughly correspond to the relativistic skin depth. Under these conditions, the laser field can penetrate the whole target and volumetrically accelerate all gold ions in the laser spot. The laser, target and simulation parameters are listed in Table 1. The laser pulse propagates in the "+x" direction and is linearly polarized in the "y" direction. The laser intensity is $sin^2$ in time and Gaussian in space, $I(t,y) = I_0 \sin^2\left(\pi t / 2\tau_{FWHM}\right) \exp\left(-(y/r_0)^2\right)$, having radius $r_0 = \frac{1}{2\sqrt{ln(2)}} D_{FWHM}$ at $1/e$ level. The laser energy is calculated according to $\varepsilon_{laser} = \pi r_0^2 I_0 \tau_{FWHM} \cong 1.13 D_{FWHM}^2 I_0 \tau_{FWHM}$. The focal spot size $D_{FWHM}$ must be carefully chosen. Additional simulations showed an increase of laser energy coupling to ions with $D_{FWHM}$ increasing, very steep for $D_{FWHM} < 5\,\mu m$, and more gentle for $D_{FWHM} > 5\,\mu m$. We adopted the value of 5 μm. Particles are initialized with charge +1 for ions and −1 for electrons. During the simulations the ion charge of oxygen and gold is dynamically incremented using a standard Monte Carlo scheme [32,33].

We focus on the most important ion beam properties, specifically charge distribution, angular distribution and flux in the forward direction. The charge distribution of Au ions, shown in Figure 4a, is generated only from ions with energy >100 MeV (>0.5 MeV/nucleon) and momentum vector within 10 degrees half-angle from the target normal, corresponding to solid angle $d\Omega = 0.095\,sr$. The maximum and average charge-to-mass ratios are 0.3 and 0.25, respectively. Optical field ionization stalls at ion charge 51 and as a result, about half of the ions pile up at $q = 51$, which corresponds to $q/M \cong 0.25$. Only a small fraction of ions with $0.25 < q/M < 0.3$ are observed. The angular distribution is highly peaked, which leads to a large flux in the forward direction. Most ions lie in a cone of ~20 degrees from the target normal (Figure 4b). There is a group of ions scattered backward, presumably from Coulomb explosion of the Au layer. According to the simplified theory of RPA, the ions located initially in the compression layer will undergo RPA and will be snow-plowed forward because for these ions the electrostatic pressure balances the radiation pressure, while the plasma containing a sheath of bare ions in the electron depletion layer will Coulomb explode launching ions in the backward direction [4]. It is interesting to note that both forward accelerated and backward scattered ions have very narrow angular distributions, i.e. both are emitted perpendicular to the foil surface. The spectra of protons and gold ions in the forward direction, $\frac{dN}{dEd\Omega}$, are plotted in Figure 5. For both protons and gold ions the cut-off energy is $E/M$>0.5 MeV/nucleon and only ions moving in a solid angle $d\Omega = 0.095\, sr$ are collected. The maximum proton energy is 85 MeV. The calculated ion fluxes and maximum energy per nucleon in the forward direction are listed in Table 2. The normalized maximum ion energy increases with $q/M$, however, this increase is closer to linear: $(E/M)_{max} \sim q/M$, rather than quadratic as it was previously found.

### 5. Heavy ion acceleration in the TNSA regime

Analogous numerical simulations are performed in the TNSA regime by increasing the foil thickness to 200 nm. The charge distribution of forward accelerated gold ion is shifted toward lower charges between 30 and 50 (Figure 6a). Now about 75 % on the ions have charges lower than the bottleneck value $q$=51. There are no ions with charges $q > 51$. However, according to Figure 1 ions with charges $51 < q < 60$ should be created in the focal spot directly by optical field ionization from the laser pulse. Figure 6b indicates that just like in the RPA regime there are two groups of counter-propagating ions, one in the forward and another in the backward direction. We looked for the "missing ions" in the backward direction. Indeed, the latter contained a group of ions with charges $q > 51$. The only plausible explanation is that ions with charges $51 < q < 60$ are created in the focal spot within one skin layer by optical field ionization from the laser pulse, but instead of being accelerated forward, are moving in the opposite direction driven by Coulomb explosion of unbalanced charges. The ions on the rear side are accelerated forward by TNSA, but the electrostatic field of the sheath is lower than the laser field, therefore the ion charge stalls at $q$=51. The spectra of protons and gold ions in the forward direction are plotted in Figure 7. The proton spectrum is nearly identical to that in Figure 5a. Protons appear to be mildly affected by target thickness variation and regime of ion acceleration. The spectrum of gold ions has a Maxwellian distribution very similar to that in Figure 5b, but the maximum energy is only 2 GeV (10 MeV/nucleon).

Comparing the two regimes of ion acceleration based on ion beam parameters alone show that RPA is the favorite, but bears a lot of similarities to TNSA (Figure 4 vs. Figure 6 and Figure 5 vs. Figure 7). A more detailed examination, however, reveals different methods of acceleration. In RPA gold ions in the skin layer are ionized to very high charges (~60), then pushed by the laser piston and form the forward-directed beam. In contrast, in the TNSA regime these ions are blown backward and the forward moving ions originate from the sheath on the rear surface.

**6. Heavy ion acceleration mechanism: energy considerations**

Though the acceleration mechanism of heavy ions can be formally attributed to the well-known ones discussed in the previous sections (RPA, BoA, TNSA, etc.), it is instructive to discuss it from a different perspective: energy absorbed by the plasma from the incoming laser pulse and how it is partitioned. The reasoning for adopting this approach is straightforward: regardless of the particular acceleration mechanism, in order to make the acceleration of heavy ions more efficient, one has to maximize the energy absorption and manipulate it by channeling more energy into the desired specie (in this case, gold ions). The energy absorption and partition is of fundamental interest and the key to ion acceleration, therefore, the objectives explored in this section of the paper center on investigating the laser energy deposition into the target. Figure 6a shows the global (integrated over the computational domain) energy balance, which at any given time reads:

$$\varepsilon^{in}(t) = \varepsilon^{field}(t) + \varepsilon^{out}(t) + \varepsilon^{kin}(t). \tag{5}$$

The electromagnetic wave energy which entered the computational domain prior to time $t$, $\varepsilon^{in}(t) = H\int_{t_0}^{t}\int_{0}^{L_y} I(y,t')dydt'$, is balanced by the electromagnetic field energy $\varepsilon^{field}(t) = \frac{H}{2}\int_{0}^{L_x}\int_{0}^{L_y}\left(\varepsilon_0\vec{E}^2(x,y) + \vec{B}^2(x,y)/\mu_0\right)dxdy$ residing in the computational domain, electromagnetic energy $\varepsilon^{out}(t) = H\int_{t_0}^{t}\oint_{L}\vec{S}(t)\cdot\vec{n}d\ell$ that left in the computational domain, and specie kinetic energy $\varepsilon^{kin}(t) = \sum_{\beta}\varepsilon_{\beta}$, summed over the kinetic energies of all computational particles β. The notation $\vec{n}$ stands for unit vector pointing outward and $\vec{S} = \frac{1}{\mu_0}\vec{E}\times\vec{B}$ is the electromagnetic energy flux (Poynting vector). The laser energy lost for optical field ionization is < 0.1 % and is not further considered in the paper. The parameter $H = \sqrt{\pi}r_0$ introduced in Ref. [34] allows for transition from energy per unit length to energy. Time $t_0 = -160$ $fs$ corresponds to the moment the laser pulse enters the computational domain at spatial position $x = 0$, and time $t = 0$ is the moment it reaches the target. For time $t \leq 0$ $\varepsilon^{field}(t) = \varepsilon^{in}(t)$, i.e. the energy entering the computational domain stays as energy of the electromagnetic field since there is no interaction with the target. At time $t = 0$ the laser bullet reaches the foil. Shortly thereafter, within 1-2 laser cycles, a hot and highly overdense plasma is formed within the target, which gradually increases to density in excess of $10^3$ times the critical electron density $n_{cr} \cong 1.8 \times 10^{21} \, cm^{-3}$. Part of the electromagnetic pulse is reflected from the plasma mirror and turns around, while the transmitted part couples energy to the plasma. As a result, for $t > 0$ $\varepsilon^{field}$ starts to decrease, while $\varepsilon^{kin}$ starts to increase. The sum of the two equals the laser energy that entered the computation domain prior to time $t$, i.e. $\varepsilon^{field}(t) + \varepsilon^{kin}(t) = \varepsilon^{in}(t)$. Later in time, at $t = 160 \, fs$, the reflected pulse going in the $-x$ direction reaches the computational domain edge ($x = 0$) and starts to leave. This is seen as a sharp increase of $\varepsilon^{out}$ and a corresponding decrease of $\varepsilon^{field}$. The peak of $\varepsilon^{out}$ can be used to estimate the reflection coefficient of the plasma, $\xi^r = \varepsilon^{out}(t_{sims}) / \varepsilon^{laser}$, while the minimum of $\varepsilon^{field}$ can be used to calculate the transmission coefficient $\xi^t = \varepsilon^{field}(t_{sims}) / \varepsilon^{laser}$. The simulations show that $\xi^r \cong 80\%$ of the laser energy is reflected and completely lost, $\xi^t \cong 5\%$ is transmitted through the target and the remaining 15 % is coupled to the plasma. The small transmission coefficient indicates that during the acceleration process the plasma remains opaque, consist with the definition for RPA. The individual terms of the energy balance are plotted in Figure 8a. Due to imperfections in the numerical discretization, Formula (5) is not exactly fulfilled. A small fraction (a few percent) of the energy "leaks" (i.e. lost) since the numerical procedure does not ensure exact energy conservation [34], unless it is artificially enforced [35]. This is acceptable, keeping in mind that the PIC simulations are computationally very intensive, but the relative error in the energy balance can be controlled by reducing the time step and/or increasing the number of computational particles [34].

Of primary interest to our investigation is the laser energy converted into specie kinetic energy. The kinetic energy increases during the pulse ($0 \leq t \leq 2\tau_{FWHM}$) and then levels off. About 4 Joules worth of laser energy is converted into kinetic energy, which is ~15 % of the laser energy on target. This energy is distributed among the species: electrons (1.9 %), gold ions from the bulk (8.3 %), protons (2.1 %) and oxygen ions (2.9 %) from the contaminant layer. Figure 8b plots the time evolution of energy absorbed by individual species. At the end of the simulations more energy has been coupled to Au compared to both oxygen and protons. At these conditions, the contaminants are no longer a problem. This is accomplished due to the appropriate choice of laser and target parameters. Next, we investigate the coupling efficiency as a function of laser intensity. As is well known, in the limiting case of low intensities the laser energy is coupled exclusively to the contaminants, more specifically, protons. In the other extreme of very high intensity, as in the example in Figure 8, the opposite happens. One can argue that there is a critical laser energy/intensity, below which the contaminants "win" and above which lays the regime suitable for heavy ion acceleration. An intensity scan can pinpoint the critical laser intensity.

Simulation results are plotted in Figure 9 for peak laser intensities between $5\times10^{20}\,W/cm^2$ and $3\times10^{21}\,W/cm^2$. The laser energy varies from 4.5 to 27 Joules. The total and individual conversion efficiencies into ions η are plotted in Figures 9a and 9b. With laser intensity increasing the conversion efficiency into protons and oxygen ions stays flat at around 2-3 %, while the conversion efficiency into gold ions increases linearly. Only at $I_0 > 2\times10^{21}\,W/cm^2$, corresponding to ~20 J of laser energy, more energy is coupled to the bulk than to the contaminants. This is the regime best suited for heavy ion acceleration. The gold ions flux $dN/d\Omega$ and normalized maximum energy $(E/M)_{max}$ versus laser intensity are plotted in Figures 9c and 9d, respectively. The ion flux sharply increases with laser intensity due to increased conversion efficiency, but then it starts to saturate when all ions in the focal spot become accelerated. The maximum energy per nucleon increases as laser intensity (and energy) squared. From Figure 9d we conclude that in order to generate gold ions with maximum normalized energy of few MeV/nucleon, the laser energy must be at least 10 Joules.

As it was pointed out in Section 3, acceleration of ions from mid- and high-Z material is inherently inefficient. The numerical simulations presented in this section indicate that there are two general approaches to produce more energetic heavy ion beams: increase the charge-to-mass ratio and improve the energy conversion efficiency. It is widely recognized that *q/M* plays a crucial role for the ion acceleration. Boosting *q/M* is therefore essential and the potential to do so has been explored. Theoretically, for gold the maximum charge-to-mass ratio is $(q/M)_{max} \cong 0.4$, provided the maximum charge is reached. In practice, however, it is lower: the average charge-to-mass ratio is only 0.25 (Figure 4a). Increasing the laser intensity from $5\times10^{20}\,W/cm^2$ to $3\times10^{21}\,W/cm^2$ did not increase appreciably *q/M*. The conclusion we drew is that regardless of the conditions, for gold ions *q/M* is limited to about 0.25. Long pulses (~1 ps) allowing for collisional ionization to take place increased *q/M* only marginally. The only viable alternative is to put more energy into the heavy ions, which was accomplished by maximizing the energy absorption with an appropriate choice of foil thickness ($\ell \cong \ell_{skin}$) and manipulating the energy distribution among species in favor of Au with a proper choice of laser intensity ($I > 2\times10^{21}\,W/cm^2$). This is the main reason to focus on the energy balance, which played central role for identifying a regime suitable for heavy ion acceleration.

**7. Conclusion**

Acceleration of heavy ions from ultrathin (<<1 μm) foil in the RPA and TNSA regimes has been investigated theoretically using a 2D PIC code for a laser system with energy of up to 27 Joules. We established that for gold ions the charge-to-mass ratio is limited to about 0.3 and the only practical approach is to improve the conversion efficiency into heavy ions by the choice of foil thickness and laser intensity. Efficient acceleration is best realized for laser pulses with energy >20 Joules focused to a spot size >5 μm at intensity $>10^{21} W/cm^2$, and ultrathin foils with thickness $\ell \cong \ell_{skin} \cong 20-30\, nm$. The laser interaction with the foil generates two collimated counter-propagating ion beams from the bulk of the foil, along the laser propagation direction and in the backward direction. The forward accelerated beam has maximum normalized energy 25 MeV/nucleon and flux $2\times10^{11}$ ions/sr.

**Acknowledgements:**

This work was performed with the support of the Air Force Office of Scientific Research under grant FA9550-14-1-0282. G. M. P. would like to acknowledge the DoD HPC computing program at NRL.

**Figure captions:**

Figure 1. Maximum charge-to-mass ratio for gold ions vs. peak laser intensity. Only optical field ionization is accounted for. Collisional ionization is neglected.

Figure 2. Maximum energy (a) and conversion efficiency into gold ions (b) vs. time. The yellow shaded area is the laser pulse profile. The laser and foil parameters are listed in Table 1.

Figure 3. A sketch of the ion acceleration mechanisms versus foil thickness.

Figure 4. Charge distribution at the end of the simulations within solid angle $d\Omega = 0.095\, sr$ (a) and angular distribution (b) of energetic (>100 MeV) gold ions in the RPA regime. The laser and foil parameters are listed in Table 1.

Figure 5. Energy spectra in the forward direction at the end of the simulations of energetic (>0.5 MeV/nucleon) gold ions (a) and protons (b) in the RPA regime. Only ions with energy within solid angle $d\Omega = 0.095\, sr$ are shown. The laser and foil parameters are listed in Table 1.

Figure 6. Charge distribution at the end of the simulations within solid angle $d\Omega = 0.095\, sr$ (a) and angular distribution (b) of energetic (>100 MeV) gold ions in the TNSA regime. The laser parameters are listed in Table 1. Foil thickness L=200 nm.

Figure 7. Energy spectra in the forward direction at the end of the simulations of energetic (>0.5 MeV/nucleon) gold ions (a) and protons (b) in the TNSA regime. Only ions with energy within solid angle $d\Omega = 0.095\, sr$ are shown. The laser parameters are listed in Table 1. Foil thickness L=200 nm.

Figure 8. (a) Energy balance components in Equation 1 versus time: energy entering the computational domain $\varepsilon^{in}$, energy leaving the computational domain $\varepsilon^{out}$, electromagnetic field energy $\varepsilon^{field}$ and kinetic energy $\varepsilon^{kin}$. Time $t_0 = -160\, fs$ corresponds to the moment the laser pulse enters the computational domain and time $t = 0\, fs$ is the moment the laser pulse reaches the target. (b) Energy absorption by electrons and ions versus time. The laser and foil parameters are listed in Table 1.

Figure 9. Total conversion efficiency into ions (a), conversion efficiency into gold ions, oxygen ions and protons (b), Au ion flux (c) and maximum energy per nucleon (d) versus laser intensity. Only ions with energy >100 MeV within 10 degrees half-angle from the target normal ($d\Omega = 0.095\, sr$) are included. The relation between laser energy and peak intensity is $\varepsilon_{laser}(J) = 9 \times I_0(W/cm^2)/10^{21}$.

Table 1. Laser, target and computational domain parameters used in the simulations.

| parameter | variable & units | value |
|---|---|---|
| laser intensity | $I_0 \, (W/cm^2)$ | $3 \times 10^{21}$ |
| pulse duration | $\tau_{FWHM} \, (\mu m)$ | 32 |
| focal spot size | $D_{FWHM} \, (\mu m)$ | 5 |
| wavelength | $\lambda \, (\mu m)$ | 0.8 |
| energy | $\varepsilon_{laser} \, (J)$ | 27 |
| foil thickness | $L \, (nm)$ | 20 |
| foil width | $W \, (\mu m)$ | 126 |
| computational domain | $L_x \times L_y \, (\mu m^2)$ | 100x128 |
| cell size | $\Delta x \times \Delta y \, (nm^2)$ | 20x20 |
| time step | $\Delta t \, (\lambda/c)$ | 0.005 |
| simulation time | $t_{sims} \, (fs)$ | 320 |

Table 2. Calculated flux, average charge-to-mass-ratio and maximum energy per nucleon in the forward direction for protons, oxygen and gold ions. Only ions with energy >100 MeV within 10 degrees half-angle from the target normal ($d\Omega = 0.095 \, sr$) are included. The laser and foil parameters are listed in Table 1.

| parameter | protons | O ions | Au ions |
|---|---|---|---|
| $dN/d\Omega$ | $2.2 \times 10^{12}$ | $3.8 \times 10^{11}$ | $1.7 \times 10^{11}$ |
| $\bar{q}/M$ | 1 | 0.5 | 0.25 |
| $(E/M)_{max}$ | 85 | 40 | 25 |

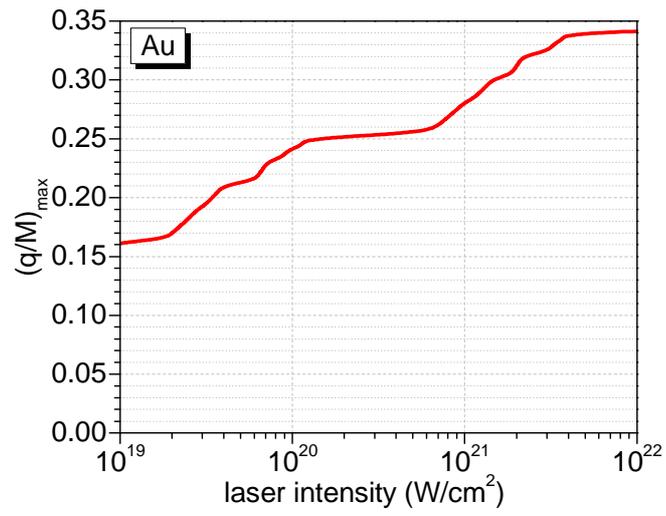

Figure 1

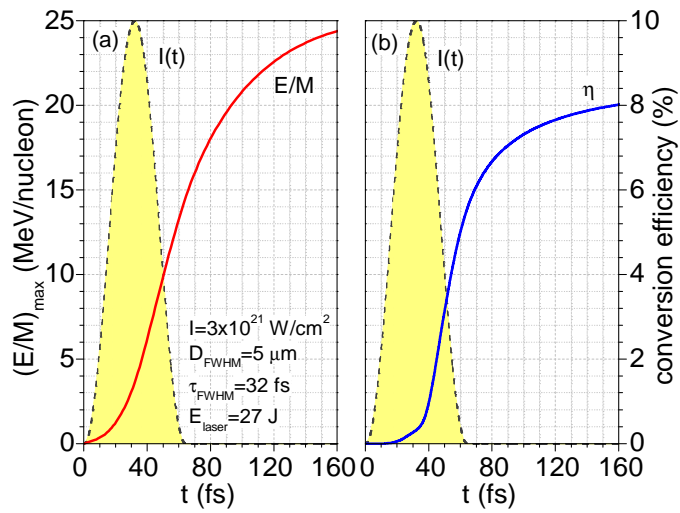

Figure 2

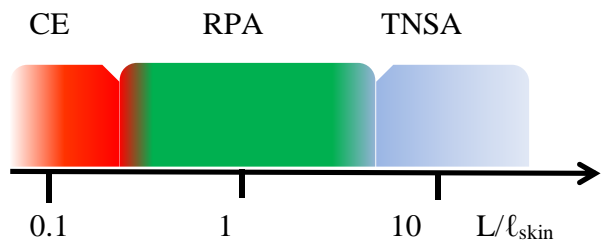

Figure 3

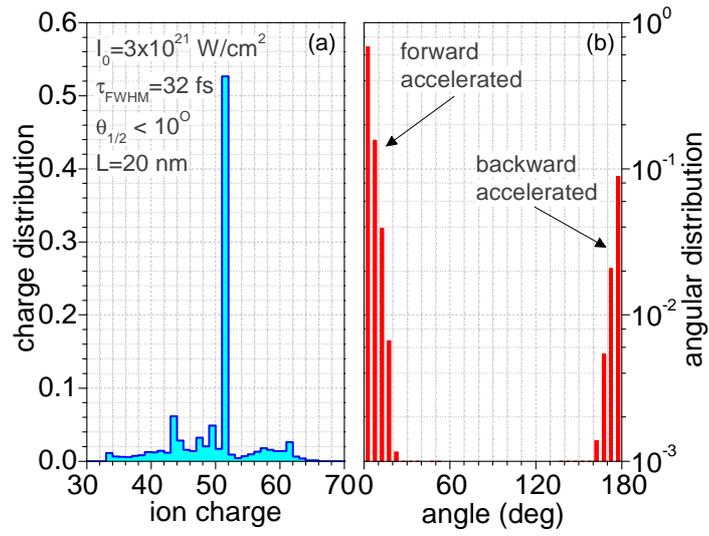

Figure 4

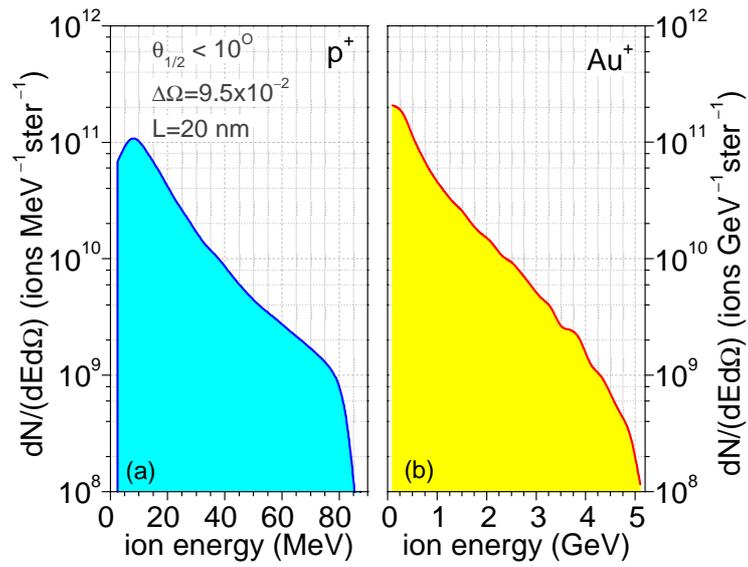

Figure 5

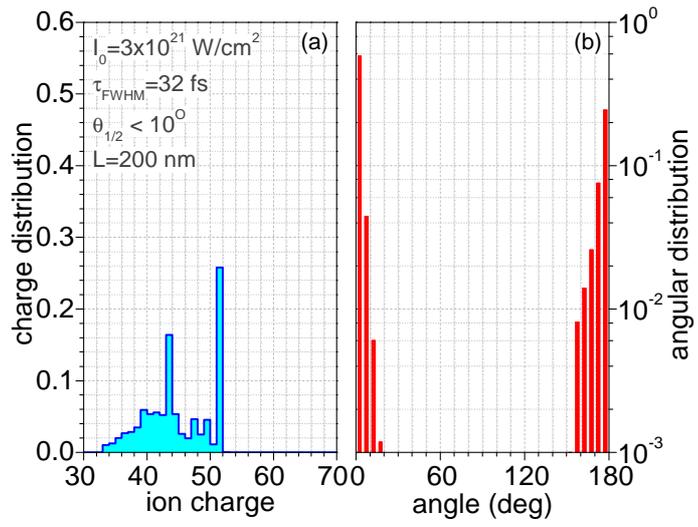

Figure 6

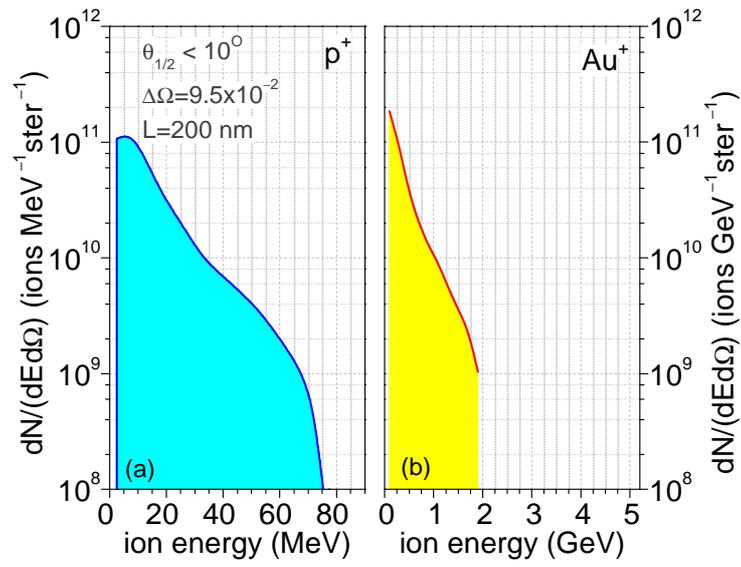

Figure 7

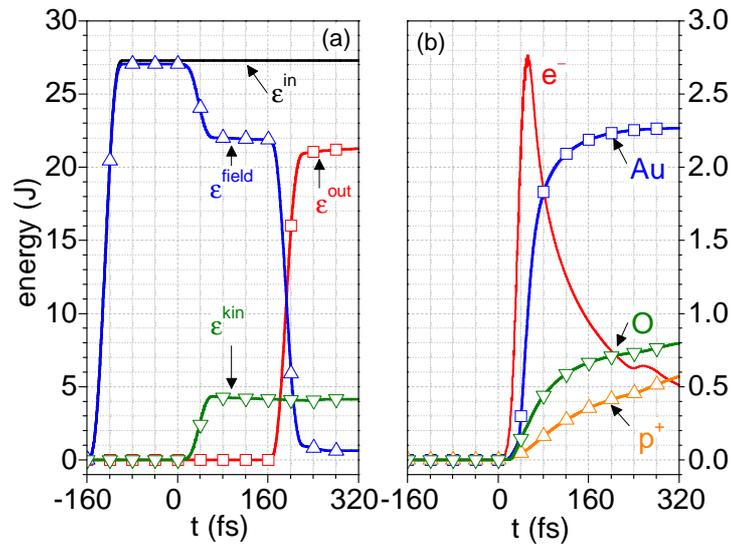

Figure 8

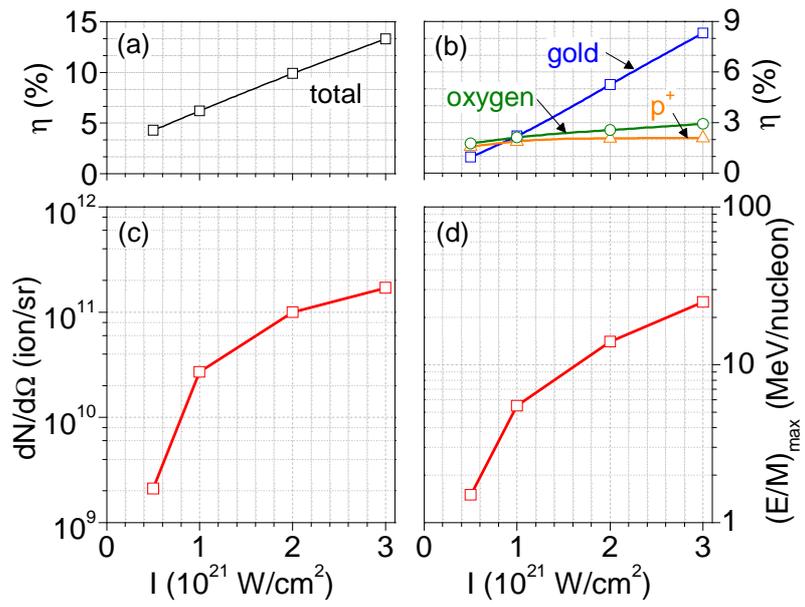

Figure 9